\documentclass[lettersize,journal]{IEEEtran}
\usepackage{amsmath,amsfonts}
\usepackage{algorithmic}
\usepackage{algorithm}
\usepackage{array}
\usepackage[caption=false,font=normalsize,labelfont=sf,textfont=sf]{subfig}
\usepackage{textcomp}
\usepackage{stfloats}
\usepackage{url}
\usepackage{verbatim}
\usepackage{graphicx}
\usepackage{cite}
\usepackage{booktabs}   
\usepackage{tabularx}
\usepackage{multirow}  
\usepackage{makecell}   
\usepackage{graphicx}   
\usepackage{adjustbox}
\usepackage{bm}
\usepackage{quantikz}
\usepackage[colorlinks, linkcolor=blue, citecolor=blue, urlcolor=blue]{hyperref}
\usepackage[capitalize]{cleveref}
\begin{document}

\title{Feature Entanglement-based Quantum Multimodal Fusion Neural Network}

\author{Yu Wu, Qianli Zhou, Jie Geng, Xinyang Deng, Wen Jiang
	\thanks{Yu Wu, is with the School of Electronics and Information, Northwestern Polytechnical University, Xi’an, 710072, China and also with the School of Computer Science and Technology, Northwestern Polytechnical University, Xi’an, 710072, China.}
	
	\thanks{Qianli Zhou, Jie Geng, Xinyang Deng, and Wen Jiang are with the School of Electronics and Information, Northwestern Polytechnical University, Xi’an, 710072, China.}
	
	\thanks{This work is partially supported by the Chinese Postdoctoral Science Foundation‌ (Grant No. 2025M784414).}

	\thanks{Manuscript received April 19, 2021; revised August 16, 2021.}}

\markboth{Journal of \LaTeX\ Class Files,~Vol.~14, No.~8, August~2021}%
{Shell \MakeLowercase{\textit{et al.}}: A Sample Article Using IEEEtran.cls for IEEE Journals}

\IEEEpubid{0000--0000/00\$00.00~\copyright~2021 IEEE}

\maketitle

\begin{abstract}
Multimodal learning aims to enhance perceptual and decision-making capabilities by integrating information from diverse sources. However, classical deep learning approaches face a critical trade-off between the high accuracy of \emph{black-box} feature-level fusion and the interpretability of less outstanding decision-level fusion, alongside the challenges of parameter explosion and complexity. This paper discusses the accuracy-interpretablity-complexity dilemma under the quantum computation framework and propose a feature entanglement-based quantum multimodal fusion neural network. The model is composed of three core components: a classical feed-forward module for unimodal processing, an interpretable quantum fusion block, and a quantum convolutional neural network (QCNN) for deep feature extraction. By leveraging the strong expressive power of quantum, we have reduced the complexity of multimodal fusion and post-processing to linear, and the fusion process also possesses the interpretability of decision-level fusion. The simulation results demonstrate that our model achieves classification accuracy comparable to classical networks with dozens of times of parameters, exhibiting notable stability and performance across multimodal image datasets. 
\end{abstract}


\section{Introduction}
\label{sec:introduction}

\IEEEPARstart{M}{ultimodal} fusion neural networks have emerged as powerful tools for enhancing perceptual and decision-making capabilities across a wide range of challenging recognition tasks. They are particularly effective in scenarios characterized by low spatial resolution or the presence of small, weak targets\cite{dong2024spectral}, where relying solely on a single data modality often fails to capture sufficient discriminative features\cite{zhao2025multidomain}. In complex environments, single-source methods encounter information bottlenecks, leading to ambiguity and vagueness. That's because physically distinct entities can appear remarkably similar within one sensory modality, yet remain clearly distinguishable in another\cite{li2022deep}. By systematically integrating these complementary sources of information, a multimodal collaborative framework can resolve such ambiguities. Therefore, constructing a multimodal framework has become a key pathway to overcoming the limitations of single-view interpretation\cite{baltruvsaitis2018multimodal}.

In the landscape of multimodal learning, fusion strategies are predominantly bifurcated into feature-level and decision-level paradigms. Feature-level fusion dominates performance benchmarks by leveraging deep neural networks to model intricate interactions. Representative transformer-based architectures, such as CLIP\cite{hafner2021clip} and BLIP\cite{li2022blip}, have established state-of-the-art standards. More specifically, incorporating texture-aware causal feature extraction \cite{xu2024texture} and cross-modal semantic enhancement mechanisms \cite{han2024cmse} can capture robust joint representations. However, these models suffer from high complexity and poor interpretability. Conversely, decision-level fusion offers a parameter-efficient and interpretable alternative\cite{chefer2021transformer}. By integrating mathematical frameworks such as Bayesian inference and Dempster–Shafer (DS) theory\cite{huang2025theory, huang2024merging}, it transforms fusion into a transparent reasoning step based on predefined rules. Although the way of integrating high-level information through established rules is highly interpretable, the effect is poor and difficult to be learned. These methods often yield lower accuracy than feature-level approaches. Therefore, a key challenge is to combine the strengths of both worlds: achieving the high precision of feature-level fusion, while preserving the interpretability and efficiency in decision-level fusion.

\IEEEpubidadjcol
Quantum computing achieves fundamentally different information processing from classical by leveraging the principles of quantum mechanics\cite{biamonte2017quantum}. Its outstanding performance in some complex problems has inspired the exploration of quantum machine learning\cite{cerezo2022challenges, shindi2023model}. With the advent of noisy intermediate-scale quantum era, research has shifted towards hybrid architectures based on variational quantum circuits (VQC) \cite{yang2024quantum, dong2019learning}. Compared to classical networks, quantum models offer distinct advantages in feature mapping and entanglement. Through feature embedding, classical data is encoded into an exponentially large Hilbert space. This enables efficient processing of non-linearly separable data in high dimensions\cite{liu2021hybrid}. Consequently, VQC-based models often require fewer parameters to achieve accuracy comparable to classical networks. This is where high expressive power and parameter efficiency lie. Furthermore, quantum entanglement captures non-local correlations within the data. This provides a physical foundation for uncovering deep relationships and enhancing model expressiveness. Therefore, we hope to transfer these advantages of quantum computing to multimodal learning.

Leveraging the aforementioned quantum advantages, quantum machine learning has rapidly expanded from unimodal tasks to the more complex domain of multimodal learning\cite{zheng2024quantum}. Recent studies demonstrate that specialized quantum fusion layers can effectively integrate heterogeneous modalities. In the field of social media sentiment classification, Li et al. \cite{li2025qmlsc} developed a model capturing subtle emotional cues better than classical counterparts. A success echoed by similar works in sarcasm and fake news detection \cite{qu2024qmfnd, phukan2024hybrid, tiwari2024quantum, singh2025quantum}. Expanding into medical diagnosis, Qu et al. \cite{qu2023qnmf} integrated diverse clinical data to improve diagnostic precision. Furthermore, for complex reasoning, Chen et al. \cite{chen2021quantum} utilized entanglement embedding for natural language question answering, and Mukesh et al. \cite{mukesh2024qvila} proposed the QVILA model, proving that quantum circuits can handle intricate vision-language interactions. However, complex quantum circuit also face significant challenges. On the one hand, most current quantum neural networks are still inherently \textit{black-box}. Their internal state evolution and feature fusion processes lack interpretability, making it difficult to quantify the information interplay between modalities. On the other hand, blindly increasing the number of qubits leads to the barren plateaus phenomenon, where gradients vanish, rendering the model untrainable\cite{larocca2025barren}. The current approach merely reduces the number of parameters without taking into account interpretability. \textbf{Therefore, the core motivation of this paper is to establish a trainable fusion method that can be theoretically explained, and it also has the advantage of parameters.}

To tackle the \textit{black-box} challenge, establishing a quantum interpretability framework with clear physical semantics has become a key approach. Quantum inference models based on DS theory have offered an approach to quantifying uncertainty. Unlike classical probability theory, this framework establishes a formal isomorphism between the DS theory structure and the quantum Hilbert space\cite{gao2025information, xiao2025adaptive}. The reason is the mathematical consistency between the power set structure and the quantum superposition state\cite{yang2016new, deng2023framework}. Thus, quantum systems can effectively model evidence combination and conflict resolution \cite{deng2023novel}. Besides, the scope of evidential reasoning has broadened to encompass both foundational theoretical extensions and diverse practical applications. Recent advancements have deepened the theoretical foundations through innovations in random permutation sets \cite{deng2024random} and the Fourier transform of basic probability assignments \cite{cheng2025fourier}. Concurrently, the framework's applicability has expanded with local differential privacy \cite{li2022local}, novel temporal fusion mechanisms \cite{zhan2025time}, and comprehensive methodologies for evidential clustering \cite{zhang2025survey, zhang2025belief}. In application domains, evidential methods have demonstrated superior efficacy in handling high-dimensional data classification \cite{gong2022sparse}, multi-source data imputation \cite{huang2024integration}, and complex decision-making in social networks \cite{wen2024eriue}. Furthermore, comprehensive reviews by Huang et al.\cite{huang2024review} highlight its role in uncertainty quantification for medical deep learning. These theoretical and practical developments provide a robust foundation for interpretable reasoning. However, this rule has not been widely applied in quantum deep learning. This paper intends to introduce this rule into multimodal learning.

To address the dual challenges of parameter explosion and lack of interpretability, this paper proposes an feature entanglement-based quantum multimodal fusion framework. It leverages the theoretical alignment between quantum computing and DS theory. Uniquely, our method transforms multimodal fusion from an opaque numerical operation into a transparent process of evidence combination. This process is governed by a conjunction introduction rule implemented through quantum entanglement. Furthermore, we introduce parameters in quantum fusion, increasing the semantic space of the importance of evidence. Consequently, this design achieves high accuracy of deep learning while ensuring logical transparency. Experiments on remote sensing benchmarks demonstrate the framework's performance and robustness. The main contributions of this research are summarized as follows: (1) Explainable quantum multimodal fusion method with clear physical semantics and logical interpretability. (2) Excellent decomposability, parallelism and scalability with extensive parameter advantage. (3) Multiple sets of tests show high accuracy and stability of our work.

The rest of this paper is outlined as follows. \cref{sec:preliminaries} introduces the preliminaries of QCNN, quantum fusion strategies and quantum evidence theory. \cref{sec:methodology} details each building block of the proposed quantum convolutional multi-modal (QCMM) framework. \cref{sec:experiments} provides the datasets, runtime environment, baselines,  experimental results and analysis, and performance comparison. Finally, \cref{sec:conclusion} concludes the paper and outlines directions for future research.

\section{Preliminaries}
\label{sec:preliminaries}

This section provides a concise overview of the key quantum machine learning concepts that form the building blocks of our QCMM framework. 

\subsection{Quantum Convolutional Neural Networks (QCNN)}
\label{subsec:qcnn}
QCNN adapts the successful hierarchical structure of classical CNNs to the quantum computing paradigm. It is designed to efficiently extract spatial or structural features from quantum data by creating a pyramidal architecture of alternately stacked layers\cite{oh2020tutorial, zhu2018quaternion}.

Quantum convolutional layer acts as the primary feature extractor. It emulates the principles of \textit{local receptive fields} and \textit{weight sharing} by applying a parameterized two-qubit unitary gate (the kernel) to adjacent qubit pairs. The $N$ width global convolutional unitary $U_c(\boldsymbol{\theta}_c)$ can be represented as:
\begin{equation}
	\label{eq:convolutional_layers}
	U_c(\boldsymbol{\theta}_c) = \left( \bigotimes_{k=1}^{N/2} u_{2k-1, 2k}(\boldsymbol{\theta}_c) \right) \left( \bigotimes_{k=0}^{N/2-1} u_{2k, 2k+1}(\boldsymbol{\theta}_c) \right)
\end{equation}
The two term corresponds to the sub-layers acting on odd-even pairs and even-odd pairs. Parameter $\boldsymbol{\theta}_c$ is shared across all local gates. The state evolution of the input density matrix $\rho_{i}$ through convolutional layer is governed by:
$
	\rho_{o} = U_c(\boldsymbol{\theta}_c) \rho_{i} U_c^{\dagger}(\boldsymbol{\theta}_c)
$.
This process entangles local qubits to capture spatial correlations within data. The specific architecture of the two-qubit kernel $u(\boldsymbol{\theta}_c)$ varies. Its design  considers the balance of entanglement, expression and learning ability.

Quantum pooling layer follows convolution. The pooling layer reduces the system's dimensionality (the number of qubits) while retaining the most significant features. A common strategy is to use parameterized controlled gates to transfer information from control qubit to target qubit. The control qubit is then discarded or traced out. 
For an input density matrix $\rho_{i}$, the pooled state $\rho_{p}$ is obtained by applying a parameterized pooling unitary $U_{p}(\boldsymbol{\theta}_p)$ and tracing out the subset of control qubits $S$:
\begin{equation}
	\rho_{p} = \text{Tr}_{S} \left( U_{p}(\boldsymbol{\theta}_p) \rho_{i} U_{p}^{\dagger}(\boldsymbol{\theta}_p) \right)
\end{equation}
This operation effectively halves the feature space size in a learnable downsampling process.

By stacking these layers, a QCNN progressively builds higher-level feature representations from the input data, making it a effective tool for quantum feature extraction.

\subsection{Quantum Multimodal Fusion Strategies}
The core objective of a quantum fusion layer is to model the complex interdependencies between different data modalities by generating targeted entanglement\cite{schuld2018supervised}. The overall process can be broken down into two key stages:

\subsubsection{Data encoding}
Data encoding maps a classical feature vector $\mathbf{x} = (x_1, \dots, x_d)$ into a quantum state $|\psi\rangle$ within the Hilbert space\cite{schuld2015introduction}. The choice of encoding strategy defines the initial quantum data structure. 

Angle encoding maps each feature $x_j$ to the rotation angle of a specific qubit, typically using $R_y$ gates. The resulting state is a separable product state:
\begin{equation}
	|\psi\rangle = \bigotimes_{j=1}^{d} R_y(x_j) |0\rangle_j
\end{equation}
Since the qubits remain uncorrelated initially, this method is well-suited for parallel, bit-wise fusion operations.

Amplitude Encoding embeds the normalized vector $\mathbf{x}$ into the probability amplitudes of an $n$-qubit system ($d = 2^n$):
\begin{equation}
	|\psi\rangle = \sum_{k=0}^{d-1} x_k |k\rangle, \quad \text{s.t. } \sum |x_k|^2 = 1
\end{equation}
While highly qubit-efficient, it creates a complex, pre-entangled state and requires deep state preparation circuits, making local feature manipulation difficult.

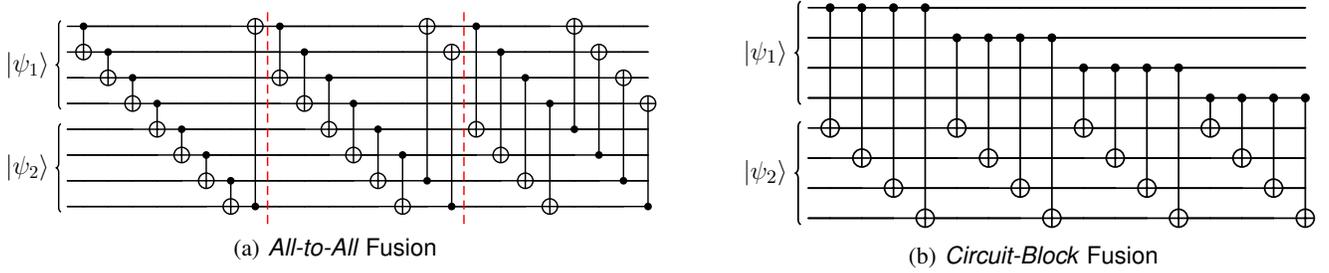
\begin{figure*}[!t]
	\centering
	\captionsetup[subfloat]{font=small, labelfont=small}
	\subfloat[\textit{All-to-All} Fusion]{
		\begin{adjustbox}{scale=0.685} 
			\begin{quantikz}[row sep={0.5cm,between origins}, column sep=0.15cm]
				\lstick[wires=4]{\fontsize{14pt}{7.5pt}\selectfont $|\psi_{1}\rangle$} & \ctrl{1} & \qw & \qw & \qw & \qw & \qw & \qw & \targ{} \slice{} & \ctrl{2} & \qw & \qw & \qw & \qw & \qw & \targ{} & \qw \slice{} & \ctrl{4} & \qw & \qw & \qw & \targ{} & \qw & \qw & \qw \\
				\qw & \targ{} & \ctrl{1} & \qw & \qw & \qw & \qw & \qw & \qw & \qw & \ctrl{2} & \qw & \qw & \qw & \qw & \qw & \targ{} & \qw & \ctrl{4} & \qw & \qw & \qw & \targ{} & \qw & \qw \\
				\qw & \qw & \targ{} & \ctrl{1} & \qw & \qw & \qw & \qw & \qw & \targ{} & \qw & \ctrl{2} & \qw & \qw & \qw & \qw & \qw & \qw & \qw & \ctrl{4} & \qw & \qw & \qw & \targ{} & \qw \\
				\qw & \qw & \qw & \targ{} & \ctrl{1} & \qw & \qw & \qw & \qw & \qw & \targ{} & \qw & \ctrl{2} & \qw & \qw & \qw & \qw & \qw & \qw & \qw & \ctrl{4} & \qw & \qw & \qw & \targ{} \\
				\lstick[wires=4]{\fontsize{14pt}{7.5pt}\selectfont $|\psi_{2}\rangle$} & \qw & \qw & \qw & \targ{} & \ctrl{1} & \qw & \qw & \qw & \qw & \qw & \targ{} & \qw & \ctrl{2} & \qw & \qw & \qw & \targ{} & \qw & \qw & \qw & \ctrl{-4} & \qw & \qw & \qw \\
				\qw & \qw & \qw & \qw & \qw & \targ{} & \ctrl{1} & \qw & \qw & \qw & \qw & \qw & \targ{} & \qw & \ctrl{2} & \qw & \qw & \qw & \targ{} & \qw & \qw & \qw & \ctrl{-4} & \qw & \qw \\
				\qw & \qw & \qw & \qw & \qw & \qw & \targ{} & \ctrl{1} & \qw & \qw & \qw & \qw & \qw & \targ{} & \qw & \ctrl{-6} & \qw & \qw & \qw & \targ{} & \qw & \qw & \qw & \ctrl{-4} & \qw \\
				\qw &\qw & \qw & \qw & \qw & \qw & \qw & \targ{} & \ctrl{-7} & \qw & \qw & \qw & \qw & \qw & \targ{} & \qw & \ctrl{-6} & \qw & \qw & \qw & \targ{} & \qw & \qw & \qw & \ctrl{-4}
			\end{quantikz}
		\end{adjustbox}
		\label{fig:fusion_ata}
	}
	\hspace{0.5cm} 
	\captionsetup[subfloat]{font=small, labelfont=small}
	\subfloat[\textit{Circuit-Block} Fusion]{
		\begin{adjustbox}{scale=0.8} 
			\begin{quantikz}[row sep={0.5cm,between origins}, column sep=0.2cm]
				\lstick[wires=4]{\fontsize{13pt}{7.5pt}\selectfont $|\psi_{1}\rangle$} & \ctrl{4} & \ctrl{5} & \ctrl{6} & \ctrl{7} & \qw & \qw & \qw & \qw & \qw & \qw & \qw & \qw & \qw & \qw & \qw & \qw \\
				\qw & \qw & \qw & \qw & \qw & \ctrl{3} & \ctrl{4} & \ctrl{5} & \ctrl{6} & \qw & \qw & \qw & \qw & \qw & \qw & \qw & \qw \\
				\qw & \qw & \qw & \qw & \qw & \qw & \qw & \qw & \qw & \ctrl{2} & \ctrl{3} & \ctrl{4} & \ctrl{5} & \qw & \qw & \qw & \qw \\
				\qw & \qw & \qw & \qw & \qw & \qw & \qw & \qw & \qw & \qw & \qw & \qw & \qw & \ctrl{1} & \ctrl{2} & \ctrl{3} & \ctrl{4} \\
				\lstick[wires=4]{\fontsize{13pt}{7.5pt}\selectfont $|\psi_{2}\rangle$} & \targ{} & \qw & \qw & \qw & \targ{} & \qw & \qw & \qw & \targ{} & \qw & \qw & \qw & \targ{} & \qw & \qw & \qw \\
				\qw & \qw & \targ{} & \qw & \qw & \qw & \targ{} & \qw & \qw & \qw & \targ{} & \qw & \qw & \qw & \targ{} & \qw & \qw \\
				\qw & \qw & \qw & \targ{} & \qw & \qw & \qw & \targ{} & \qw & \qw & \qw & \targ{} & \qw & \qw & \qw & \targ{} & \qw \\
				\qw & \qw & \qw & \qw & \targ{} & \qw & \qw & \qw & \targ{} & \qw & \qw & \qw & \targ{} & \qw & \qw & \qw & \targ{}
			\end{quantikz}
		\end{adjustbox}
		\label{fig:fusion_cb}
	}
	
	\caption{Different topology of entangling fusion circuit: (a) \textit{All-to-All} and (b) \textit{Circuit-Block}. $|\psi_{1}\rangle$ and $|\psi_{2}\rangle$ represent different modalities.}
	\label{fig:baseline_fusions}
\end{figure*}

\subsubsection{Quantum fusion strategies}
The quantum fusion process can be summarized as follows. The quantum states representing different modalities are brought together, and an entangling circuit is applied to create cross-modal correlations. Research has focused on the topology of fusion circuit recently. Some prominent structures are shown in fig. \ref{fig:baseline_fusions}:
\hypertarget{target:fusion-strategy}{\textit{All-to-all}}\cite{shende2004minimal}
forms a fully connected graph, maximizing the potential for capturing global correlations but at the cost of significant circuit depth and susceptibility to noise. \textit{Circuit-block}\cite{zheng2024quantum} uses structured, repeating patterns of gates to offer a practical balance between entangling capability and trainability on near-term quantum devices.

Different fusion circuits make trade-offs in hardware efficiency, that is, balancing circuit depth, the number of gates, flexibility, learning ability, and entanglement ability. Beyond these structural designs, recent approaches have begun to explore how to imbue the fusion process itself with clearer logical or physical semantics, moving beyond \textit{black-box} entanglement towards more interpretable frameworks.

\subsection{Quantum Evidence Theory and Fusion Implementation}
\label{subsec:quantum_evidence}

Quantum evidence theory has emerged as a framework to bridge the gap between \textit{black-box} computation and logical interpretability\cite{gao2025quantum, xiao2022negation}. By establishing a mathematical isomorphism between DS theory and quantum mechanics, this framework enables evidential reasoning to be performed directly on quantum circuits.

\subsubsection{Mass function and evidence state}
In classical DS theory, the \textit{frame of discernment} $\Omega = \{\omega_1, \dots, \omega_C\}$ represents the set of mutually exclusive hypotheses. And $2^{\Omega} = \{\emptyset, \{\omega_1\}, \{\omega_2\}, \{\omega_1, \omega_2\}, \ldots, \Omega\}$ donates its power set. A \textit{basic probability assignment}, or mass function $m$, assigns a belief value to each subset $A \subseteq \Omega$, satisfying $\sum_{A} m(A) = 1$.

Existing studies have formalized the mapping of this power set structure onto the Hilbert space. A quantum evidence state $|\mathcal{M}\rangle$ is defined where orthogonal basis states represent the subsets of $\Omega$, and probability amplitudes encode the belief masses. Formally, for a mass function $m$, it's expressed as 
$
	|\mathcal{M}\rangle = \sum_{A \subseteq \Omega} \sqrt{m(A)} e^{i\phi_A} |A\rangle
	\label{eq:qes}
$, 
where $|A\rangle$ is the basis state corresponding to the element $A$ in $2^{\Omega}$. And the phase $\phi_A$ denotes the phase angle.

\subsubsection{Quantum fusion strategy}
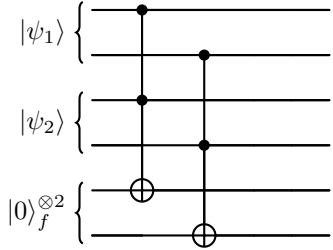
\begin{figure}[h!]
	\centering
	\begin{quantikz}[row sep={0.6cm,between origins}]
		\lstick[wires=2]{$|\psi_{1}\rangle$} & \ctrl{4} & \qw & \qw & \qw & \qw \\
		& \qw & \ctrl{4} & \qw & \qw & \qw \\
		\lstick[wires=2]{$|\psi_{2}\rangle$} & \ctrl{2} & \qw & \qw & \qw & \qw \\
		& \qw & \ctrl{2} & \qw & \qw & \qw \\
		\lstick[wires=2]{$|0\rangle^{\otimes 2}_f$} & \targ{} & \qw & \qw & \qw & \qw \\
		& \qw & \targ{} & \qw & \qw & \qw
	\end{quantikz}
	\caption{A quantum evidential fusion circuit for two modalities with 2 qubits.}
	\label{fig:fusion_circuit}
\end{figure}

The core of evidential fusion is the \textit{conjunctive combination rule} in DS theory. For two independent mass functions $m_1$ and $m_2$, the combined mass $m(C)$ for $C \subseteq \Omega$ is proportional to the orthogonal sum of their intersection:
\begin{equation}\label{eq:dempster_rule}
	m(C) = \frac{1}{1 - K} \sum_{A \cap B = C} m_1(A) m_2(B)
\end{equation}
where $K = \sum_{A \cap B = \emptyset} m_1(A) m_2(B)$
is the \emph{conflict coefficient}. In the quantum circuit, this mathematical intersection logic ($A \cap B$) can be physically realized through Toffoli gate. As illustrated in Fig. \ref{fig:fusion_circuit}, the target qubits flip \textit{if and only if} the control qubits from different modalities are simultaneously in the active state $|1\rangle$. The resulting target amplitude corresponds to the product of input beliefs. Thus, it physically simulates the mathematical conjunction $m_1(A)m_2(B)$.

\section{Methodology}
\label{sec:methodology}

\subsection{Problem Formalization and Motivation}

\subsubsection{Motivation}
As highlighted in \cref{sec:introduction}, in order to achieve both the effect of feature-level fusion and the interpretability of decision fusion, we turn to the unique properties of quantum mechanics, which offers a fundamentally different approach to data integration. We identify a mathematical isomorphism between this quantum state evolution and the logic of DS theory. The probabilistic nature of quantum amplitudes and its distribution naturally aligns with the evidence accumulation process.

Furthermore, quantum networks can physically bind distinct data sources into a unified system by entanglement. Unlike classical methods that approximate feature interactions through layers of massive weight matrices, quantum evolution maps features into an inseparable joint state within an exponentially large Hilbert space. This enables the network to process information holistically. Besides, the module can exploit the high-dimensional feature space to capture complex, non-linear correlations while maintaining a significant parameter advantage. This motivates our design: to leverage quantum neural networks for efficient feature fusion and extracting, while grounding the fusing logic in DS theory to resolve the interpretability crisis.

\subsubsection{Problem formalization}

This study addresses a multimodal land cover classification task. We define the dataset as a collection of $N$ samples, $\mathcal{D} = \{(\mathbf{X}_h^{(i)}, \mathbf{X}_l^{(i)}, y^{(i)})\}_{i=1}^{N}$, where $(\mathbf{X}_h^{(i)}, \mathbf{X}_l^{(i)})$ represents a pair of co-registered data patches from two distinct modalities (e.g., HSI and LiDAR), and $y^{(i)}$ is the corresponding ground-truth label belonging to a discrete set of $C$ classes. Our objective is to learn a parameterized mapping function, $F$, that accepts a multimodal input pair and predicts a probability distribution over the $C$ classes, denoted as $\hat{\mathbf{y}}^{(i)} = F(\mathbf{X}_h^{(i)}, \mathbf{X}_l^{(i)}; \boldsymbol{\Theta})$. In this work, the function $F$ is realized by the proposed QCMM framework, and $\boldsymbol{\Theta}$ represents the comprehensive set of all trainable parameters within it. The learning process involves optimizing $\boldsymbol{\Theta}$ to minimize a loss function $\mathcal{L}$ that quantifies the discrepancy between the predicted probabilities $\hat{\mathbf{y}}^{(i)}$ and the true labels $y^{(i)}$ across the training set.

\begin{figure*}[!t]
	\centering
	\includegraphics[width=\textwidth]{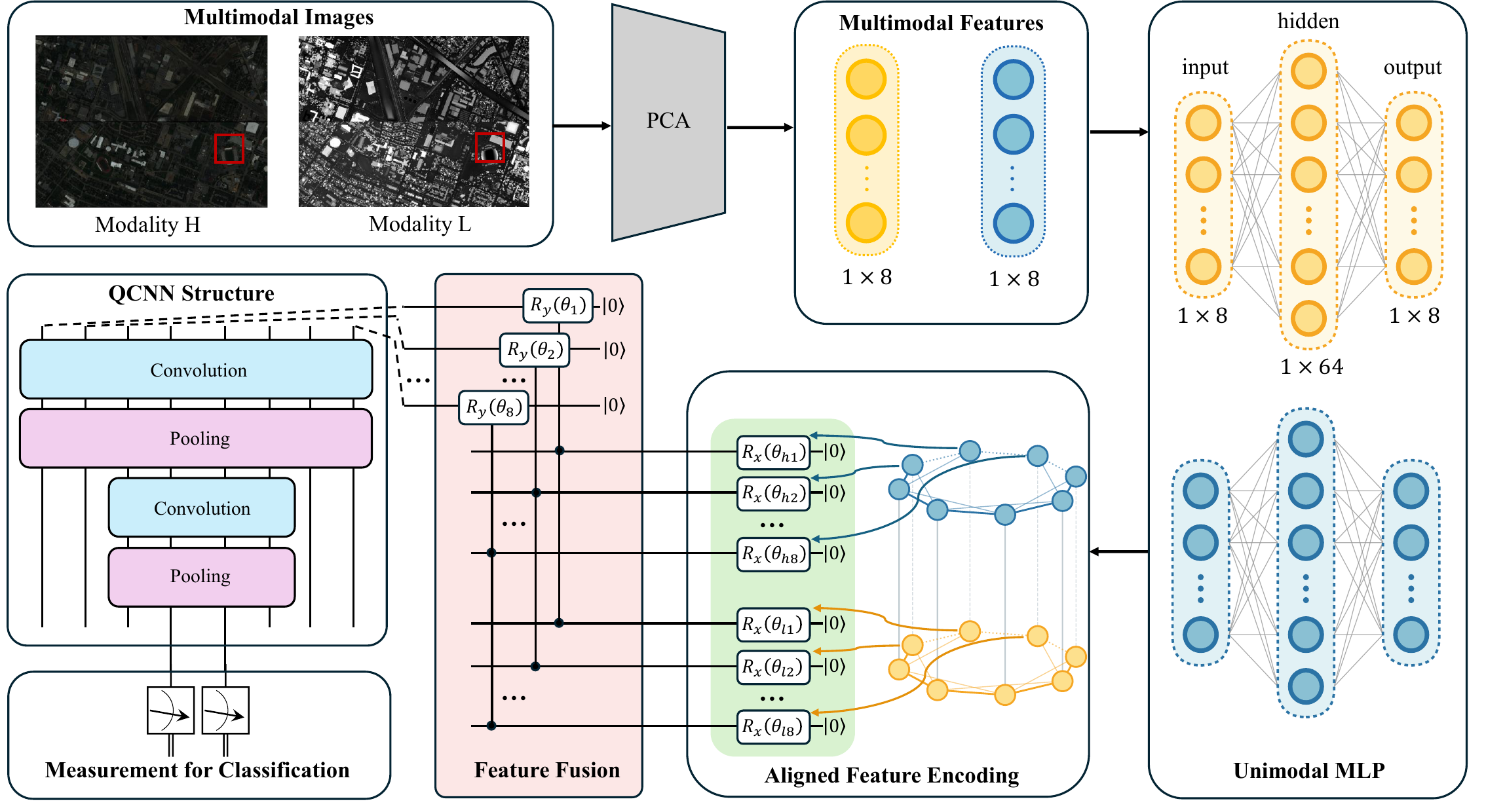}
	\caption{The overall architecture of the proposed QCMM framework. The pipeline consists of three key stages: (1) Unimodal feature alignment: Classical MLPs extract and align features. (2) Quantum embedding and fusion: Bit-wise angle encoding and trainable evidence fusion; (3) QCNN: Deep semantic extraction and compression by convolution and pooling.}
	\label{fig:qcmm_structure}
\end{figure*}

\subsection{Overall Architecture}
The proposed QCMM framework is constructed as a hybrid quantum-classical architecture. As illustrated in \cref{fig:qcmm_structure}, the framework processes a pair of co-registered data patches from two modalities through a multi-stage pipeline that is trained end-to-end. The data flow proceeds as follows:

\subsubsection{Data preprocessing} The initial raw, high-dimensional input patches, denote as $\mathbf{X}_h$ and $\mathbf{X}_l$. Principal component analysis (PCA) is used to project the data into a lower-dimensional space of dimension $d$, expressed as $\mathbf{x}_h, \mathbf{x}_l \in \mathbb{R}^d$.

\subsubsection{Unimodal feature extraction and alignment} Preprocessed data are fed into separate unimodal networks, $\mathcal{M}_h$ and $\mathcal{M}_l$. It's expressed as $\mathbf{v}_m = \mathcal{M}_m(\mathbf{x}_m), \text{for } m \in \{h, l\}$. This step extracts higher-level features while implicitly learning feature alignment for the subsequent quantum fusion.

\subsubsection{Quantum embedding and initial state preparation} Prepare the following quantum states:   $|\Psi_0\rangle = |\psi_h\rangle \otimes |\psi_l\rangle \otimes |0\rangle^{\otimes d}_f$. where $|\psi_h\rangle$ and $|\psi_l\rangle$ are the encoded states for the HSI and LiDAR modalities, $|0\rangle^{\otimes d}_f$ is the ground state of the fusion register, and $\otimes$ is the tensor cross product.

\subsubsection{Quantum fusion} A parameterized quantum fusion layer, expressed as $U_f(\boldsymbol{\theta})$ has trainable angles $\boldsymbol{\theta}$. Applying to the initial state, this operation generates entanglement within multiple modalities and fusion targets, resulting in a fused quantum state $|\Psi_f\rangle = U_f(\boldsymbol{\theta}) |\Psi_0\rangle$.

\subsubsection{Quantum deep feature extraction} The fused state $|\Psi_f\rangle$ is then processed by QCNN, expressed as $U(\boldsymbol{\phi})$, where $\boldsymbol{\phi}$ represents trainable parameters. It distills high-level semantic features, resulting as $|\Psi_i\rangle = U(\boldsymbol{\phi}) |\Psi_f\rangle$.

\subsubsection{Measurement and optimization} The state $|\Psi_i\rangle$ is measured to obtain a classical probability distribution, $\hat{\mathbf{y}}$, over the $C$ target classes. The probability for the $k$-th class is given by the Born rule, $\hat{y}_k = |\langle k | \Psi_i \rangle|^2$. The model's complete set of trainable parameters, $\boldsymbol{\Theta} = \{\mathbf{W}_h, \mathbf{W}_l, \boldsymbol{\theta}, \boldsymbol{\phi}\}$, is optimized by minimizing the loss \cref{eq:loss}.

\subsection{Quantum Multimodal Fusion Network}

\subsubsection{Data preprocessing} It works as offline dimensionality. The input dataset is composed of multimodal data tuples $\{(\mathbf{X}_{h}^{(i)}, \mathbf{X}_{l}^{(i)}, y^{(i)})\}_{i=0}^{N-1}$, where $\mathbf{X}_{h}^{(i)} \in \mathbb{R}^{S \times S \times B}$ represents the Hyperspectral Imagery (HSI) patch with $B$ spectral bands, and $\mathbf{X}_{l}^{(i)} \in \mathbb{R}^{S \times S}$ denotes the corresponding LiDAR patch. The spatial dimension is set to $S=7$. In this stage, we employ PCA to reduce the dimensionality of both modalities to $d=8$.

PCA serves as a robust linear pre-processing step to compress  high-dimensional raw data into a compact feature space suitable. By significantly reducing dimensionality while preserving dominant feature information, it effectively adapts the inputs to the constraints of current NISQ hardware. Moreover, the selection of this  standard, computationally efficient method, as opposed to complex non-linear extractors, ensures that the subsequent classification performance reflects the intrinsic ability of our framework.

\subsubsection{Unimodal feature extraction and alignment} Following the offline dimensionality reduction, the data is processed by a trainable unimodal single-layer MLP whose size of hidden layer is 64, and dimension of both input and output layers are $d=8$. This block serves two critical objectives: Firstly, it maps the preprocessed features into a non-linear latent space, enhancing the representational power of single-modality features before fusion. Most importantly, it can complete semantic feature alignment implicitly.   The unique topology of our downstream quantum circuit endows us with the supervisory capability of cross-modal alignment. Since the subsequent quantum fusion layer utilizes a one-to-one qubit interaction strategy (i.e., bit-wise controlled gates), the gradients back-propagated during end-to-end training essentially force the MLPs to adjust their outputs. This implicitly guides the networks to map corresponding semantic information from different modalities to aligned positions in the feature vectors as well as the corresponding qubits.

The transformation for each modality $m \in \{h, l\}$ is formulated as:
\begin{equation}
	\mathbf{v}_m = \mathbf{W}_2^{(m)} \sigma(\mathbf{W}_1^{(m)} \mathbf{x}_m + \mathbf{b}_1^{(m)}) + \mathbf{b}_2^{(m)}
\end{equation}
where $\mathbf{x}_m \in \mathbb{R}^d$ is the vector for modality $m$ after PCA. $\mathbf{W}_1^{(m)} \in \mathbb{R}^{k \times d}$ and $\mathbf{b}_1^{(m)} \in \mathbb{R}^k$ are the learnable weight and bias of the hidden layer, and $k=64$ is its size. $\sigma$ represents a non-linear activation function (i.e., ReLU). $\mathbf{W}_2^{(m)} \in \mathbb{R}^{d \times k}$ and $\mathbf{b}_2^{(m)} \in \mathbb{R}^d$ are the weight and bias of the output layer. $\mathbf{v}_m \in \mathbb{R}^d$ is the final aligned feature vector for the modality, which serves as the input to the quantum embedding layer.

\begin{figure*}[!t]
	\centering
	\includegraphics[width=\textwidth, trim={0.5cm 1cm 0.5cm 0cm}, clip]{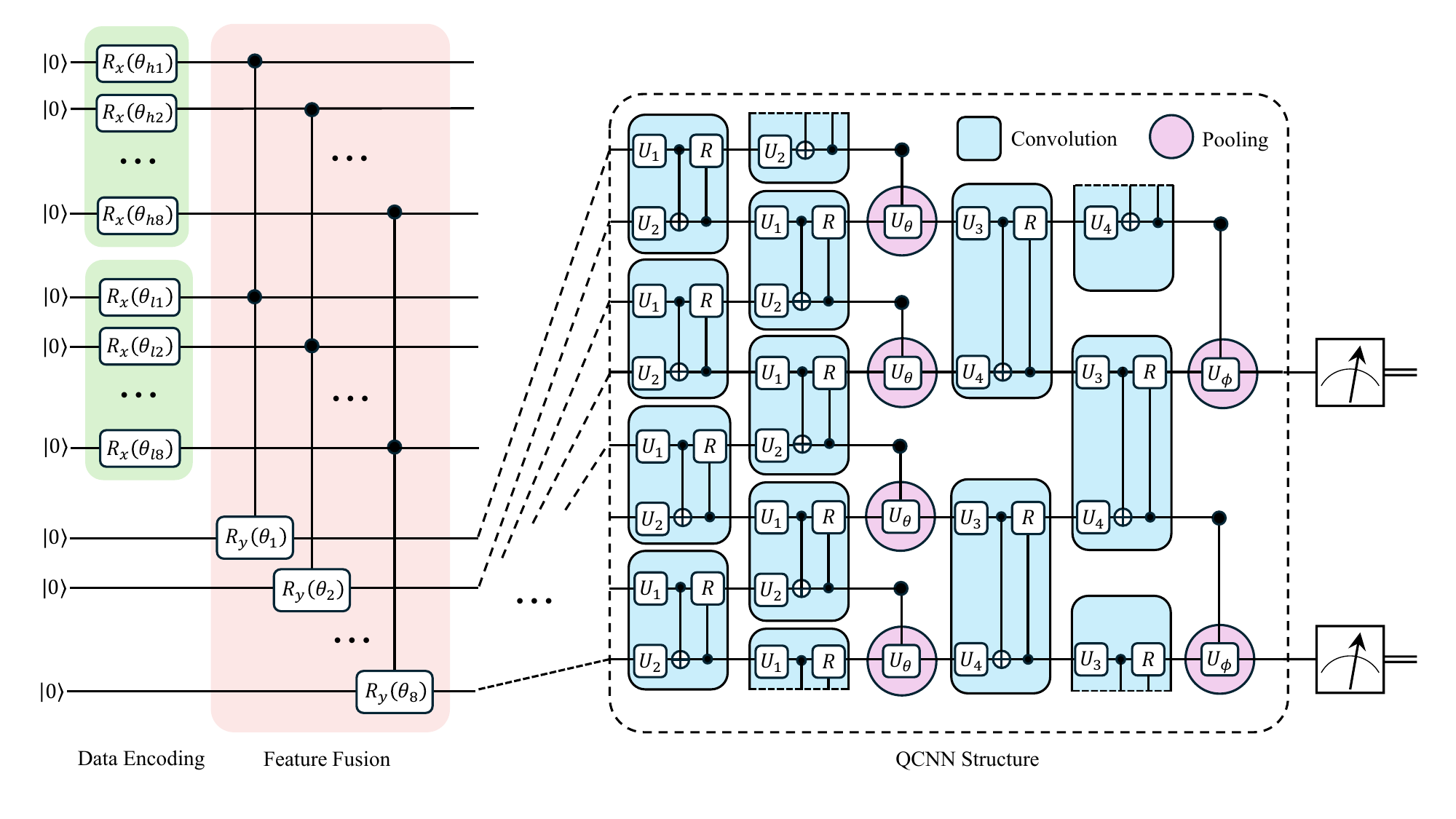}
	\caption{Detailed quantum circuit architecture of the QCMM framework. The left part shows the angle encoding and the bit-wise feature fusion. The right part details the hierarchical structure of the QCNN, which consists of two stacked convolutional-pooling layers, reducing the system from 8 qubits to 2 qubits for final measurement. The convolution and pooling kernels in the figure are just illustrative.}
	\label{fig:qcmm_overview}
\end{figure*}

\subsubsection{Quantum embedding and initial state preparation} This step will obtain three sets of qubits, namely the feature registers encoding two modalities' data respectively and the target registers to be fused with the initial state $|0\rangle^{\otimes d}$. In this phase, the aligned classical feature vectors $\mathbf{v}_h, \mathbf{v}_l \in \mathbb{R}^d$ (where $d=8$) are embedded to feature registers, denoted as $Q_h$ and $Q_l$. And the target register $Q_f$ is initialized to the ground state $|0\rangle^{\otimes d}$. We employ angle encoding to embed aligned features. Specifically, for each modality $m \in \{h, l\}$, the $j$-th component of the feature vector $v_{m,j}$ parameterizes a Rotation-$Y$ gate ($R_y$). And this $R_y$ is applied to the qubit of grand state $|0\rangle$. The encoded quantum state, $|\psi_m\rangle$, is formulated as:
$
|\psi_m\rangle = \bigotimes_{j=1}^{d} R_y(v_{m,j}) |0\rangle_j
$. 
The total initial state of the coupled $3d$-qubit system, denoted as $|\Psi_0\rangle$, expressed as the tensor product of three registers:
$
|\Psi_0\rangle = |\psi_h\rangle \otimes |\psi_l\rangle \otimes |0\rangle^{\otimes d}_f
$.

\subsubsection{Quantum fusion} \hypertarget{target:fusion}{} This layer executes the core fusion operation by applying entanglement evolution between the prepared quantum states. We implement a bit-wise interaction strategy. Parameterized unitary operator $U_f(\boldsymbol{\theta})$ is applied to the initial state $|\Psi_0\rangle$ where $\boldsymbol{\theta}$ represent the trainable parameters with the length of $d$. This operator can be decomposed into $d$ parallel local gates, with the $j$-th gate acting exclusively on the corresponding triplet of qubits $\{q_{h,j}, q_{l,j}, q_{f,j}\}$ that located at the same index across three registers. The single interaction is realized via a parameterized Controlled-Controlled-Rotation ($CC\text{-}R_y$) gate. The rotation of the target qubit $q_{f,j}$ is activated strictly conditional on the simultaneous $|1\rangle$ state of the two control qubits from the feature registers. Formulaically, The unitary operation for the $j$-th triplet, $U_f^{(j)}(\theta_j)$, is defined as:
\begin{equation}
U_f^{(j)}(\theta_j) = \left( I_{hl} - |11\rangle\langle 11|_{hl} \right) \otimes I_f + |11\rangle\langle 11|_{hl} \otimes R_y(\theta_j)_f
\end{equation}
where $\theta_j$ is a trainable parameter governing the rotation angle. And $|11\rangle\langle 11|_{hl}$ is the projection operator onto the state where both control qubits are $|1\rangle$.

Following the unitary evolution, the control registers $Q_h$ and $Q_l$ are traced out to obtain the state of the target register. Due to the bit-wise independence of the gates, the final fused density matrix $\rho_f$ can be expressed as the tensor product of the reduced density matrices from each local triplet:
\begin{equation}
	\begin{aligned}
		\rho_f = \bigotimes_{j=1}^{d} \text{Tr}_{h,l} \Bigl( 
		& U_f^{(j)}(\theta_j) \bigl( |\psi_{h,j}\rangle \langle \psi_{h,j}| \\
		& \otimes |\psi_{l,j}\rangle \langle \psi_{l,j}| \otimes |0\rangle \langle 0|_{f,j} \bigr) U_f^{(j)\dagger}(\theta_j) 
		\Bigr)
	\end{aligned}
\end{equation}
This density matrix $\rho_f$ encapsulates the fused multimodal features and serves as the input to the subsequent QCNN. And $U_f^{(j)\dagger}(\theta_j)$ represents the conjugate transpose of the unitary operation for the $j$-th triplet.

\subsubsection{QCNN} 
The QCNN module functions as the backend classifier to hierarchically distill features from the fused quantum state $\rho_f$. In our specific implementation, we construct a pyramidal architecture comprising two sequential convolutional-pooling blocks. The first block operates on the full 8-qubit register, compressing it to 4 qubits, while the second block further reduces the system to 2 qubits for final classification.

To the convolutional core, we employ a translationally invariant ansatz with weight sharing. Each global convolutional unitary $U_c(\boldsymbol{\theta}_c)$ is configured by stacking two sub-layers of local two-qubit unitaries $u(\boldsymbol{\theta}_c)$. The formulation has been given in \cref{eq:convolutional_layers}.
The value of input size $N$ is either 8 or 4. After each convolutional layer, state
$
	\rho_{o} = U_c(\boldsymbol{\theta}_c) \rho_{i} U_c^{\dagger}(\boldsymbol{\theta}_c)
$. 
We investigate three representative ansatz architectures for the local kernel $u(\boldsymbol{\theta}_c)$, as depicted in the benchmark study by Hur et al. \cite{hur2022quantum}. They are shown at fig. \ref{fig:kernel_ansatzes}:

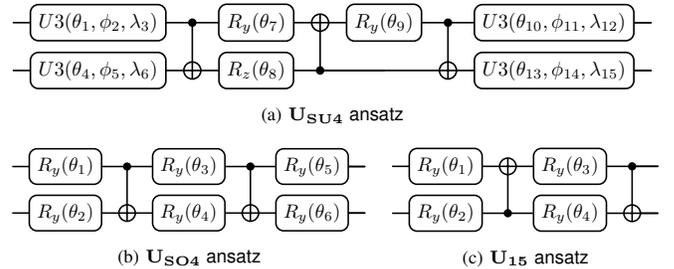
\begin{figure}[t]
	\centering
	
	\captionsetup[subfloat]{font=scriptsize, labelfont=scriptsize}
	\subfloat[$\mathbf{U_{SU4}}$ ansatz]{
		\begin{adjustbox}{width=0.99\linewidth} 
			\begin{quantikz}[row sep=0.2cm, column sep=0.3cm]
				\lstick{} & \gate[style={rounded corners}]{U3(\theta_1, \phi_2, \lambda_3)} & \ctrl{1} & \gate[style={rounded corners}]{R_y(\theta_7)} & \targ{} & \gate[style={rounded corners}]{R_y(\theta_9)} & \ctrl{1} & \gate[style={rounded corners}]{U3(\theta_{10}, \phi_{11}, \lambda_{12})} & \qw \\
				\lstick{} & \gate[style={rounded corners}]{U3(\theta_4, \phi_5, \lambda_6)} & \targ{}  & \gate[style={rounded corners}]{R_z(\theta_8)} & \ctrl{-1} & \qw                  & \targ{}  & \gate[style={rounded corners}]{U3(\theta_{13}, \phi_{14}, \lambda_{15})} & \qw
			\end{quantikz}
		\end{adjustbox}
		\label{fig:ansu4}
	}
	
	\vspace{0.2cm} 
	
	\begin{minipage}{1.0\linewidth} 
		\centering
		
		\begin{minipage}{0.56\linewidth}
			\centering
			\captionsetup[subfloat]{font=scriptsize, labelfont=scriptsize}
			\subfloat[$\mathbf{U_{SO4}}$ ansatz]{
				\begin{adjustbox}{width=\linewidth} 
					\begin{quantikz}[row sep=0.2cm, column sep=0.3cm]
						\lstick{} & \gate[style={rounded corners}]{R_y(\theta_1)} & \ctrl{1} & \gate[style={rounded corners}]{R_y(\theta_3)} & \ctrl{1} & \gate[style={rounded corners}]{R_y(\theta_5)} & \qw \\
						\lstick{} & \gate[style={rounded corners}]{R_y(\theta_2)} & \targ{}  & \gate[style={rounded corners}]{R_y(\theta_4)} & \targ{}  & \gate[style={rounded corners}]{R_y(\theta_6)} & \qw
					\end{quantikz}
				\end{adjustbox}
				\label{fig:anso4}
			}
		\end{minipage}
		\hfill 
		\begin{minipage}{0.43\linewidth}
			\centering
			\captionsetup[subfloat]{font=scriptsize, labelfont=scriptsize}
			\subfloat[$\mathbf{U_{15}}$ ansatz]{
				\begin{adjustbox}{width=\linewidth} 
					\begin{quantikz}[row sep=0.2cm, column sep=0.3cm]
						\lstick{} & \gate[style={rounded corners}]{R_y(\theta_1)} & \targ{}  & \gate[style={rounded corners}]{R_y(\theta_3)} & \ctrl{1} & \qw \\
						\lstick{} & \gate[style={rounded corners}]{R_y(\theta_2)} & \ctrl{-1} & \gate[style={rounded corners}]{R_y(\theta_4)} & \targ{}  & \qw
					\end{quantikz}
				\end{adjustbox}
				\label{fig:ans15}
			}
		\end{minipage}
	\end{minipage}
	
	\caption{Architectures of the three representative ansatzes for the local two-qubit convolution kernels.}
	\label{fig:kernel_ansatzes}
\end{figure}

\textbf{$\mathbf{U_{SO4}}$:} Designed to implement an arbitrary gate from the special orthogonal group $SO(4)$. It consists of parameterized single-qubit $R_y$ and $R_z$ rotations interlaced with CNOT gates. This ansatz is particularly suitable for tasks where the relevant information can be encoded in real-valued amplitudes, balancing expressibility with a moderate parameter count\cite{parrish2019quantum}.
	
\textbf{$\mathbf{U_{SU4}}$:} Represents the most general two-qubit unitary operation, capable of spanning the full special unitary group $SU(4)$\cite{maccormack2022branching}. It typically requires 15 trainable parameters to realize arbitrary entanglement and rotation. While computationally more expensive, it offers the theoretical maximum expressibility for a local filter.
	
\textbf{$\mathbf{U_{15}}$:} A hardware-efficient ansatz (referencing Circuit 15 from Sim et al. \cite{sim2019expressibility}) characterized by high entangling capability. It employs a deeper stack of $R_y$ rotations and CNOT gates compared to simpler circuits, designed to capture complex correlations with fewer parameters than the full $SU(4)$ ansatz.

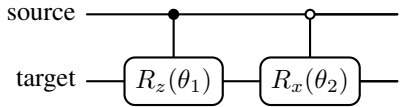
\begin{figure}[H]
	\centering
	\begin{quantikz}
		\lstick{\text{source}} & \ctrl{1} & \octrl{1} & \qw \\
		\lstick{\text{target}} & \gate[style={rounded corners}]{R_z(\theta_1)} & \gate[style={rounded corners}]{R_x(\theta_2)} & \qw
	\end{quantikz}
	\caption{The pooling layer applies two controlled rotations, $R_z(\theta_1)$ and $R_x(\theta_2)$, to compress information from the source qubit to the target qubit.}
	\label{fig:pooling}
\end{figure}

To the pooling layer, we adopt the parameterized quantum pooling circuit shown in Fig. \ref{fig:pooling} (referencing the structure in \cite{hur2022quantum}). The pooling unit operates on a source qubit $s$ and a target qubit $t$. It employs a  strategy using two controlled rotations. The local unitary $u_{s,t}(\boldsymbol{\theta}_p)$ is defined as:
\begin{equation}
	u_{s,t}(\boldsymbol{\theta}_p) = |0\rangle\langle 0|_s \otimes R_x(\vartheta_2)_t + |1\rangle\langle 1|_s \otimes R_z(\vartheta_1)_t
\end{equation}
where $\boldsymbol{\theta}_p = \{\vartheta_1, \vartheta_2\}$ are trainable parameters. Physically, this gate performs a conditional rotation on the target qubit: rotating around the X-axis by $\vartheta_2$ if the source is $|0\rangle$, and around the Z-axis by $\vartheta_1$ if the source is $|1\rangle$. The source qubit is subsequently traced out, effectively compressing the feature information into the target qubit.

\subsubsection{Measurement and optimization} We perform projective measurement on the final 2 qubits to get classification probabilities directly, in the computational basis $\mathcal{B} = \{|00\rangle, |01\rangle, |10\rangle, |11\rangle\}$. Each basis state corresponds to one of the $C=4$ land cover classes. The predicted probability $\hat{y}_{k}^{(i)}$ that the $i$-th sample belongs to class $k$ (where $k \in \{0, 1, 2, 3\}$) is quantified by the expectation value of the projection operator $P_k = |k\rangle\langle k|$. According to the Born rule, this is expressed as:
\begin{equation}
\hat{y}_{k}^{(i)} = \text{Tr}(P_k \rho_{out}^{(i)})
\end{equation}
where $\text{Tr}(\cdot)$ represents the trace operation. Since the basis states form a complete set ($\sum P_k = I$), the resulting probabilities satisfy the normalization condition $\sum_{k} \hat{y}_{k}^{(i)} = 1$.

we employ the categorical cross-entropy loss function to evaluate the discrepancy between the predicted probability distribution $\hat{\mathbf{y}}^{(i)}$ and the one-hot encoded ground truth vector $\mathbf{y}^{(i)}$. The global objective function $\mathcal{L}$ over a batch of $N$ samples is formulated in equation:
\begin{equation}
	\mathcal{L}(\boldsymbol{\Theta}) = - \frac{1}{N} \sum_{i=1}^{N} \sum_{k=0}^{C-1} y_{k}^{(i)} \log \left( \hat{y}_{k}^{(i)} \right)
	\label{eq:loss} 
\end{equation}
where $N$ is the batch size. $\boldsymbol{\Theta} = \{\mathbf{W}^{(m)}, \boldsymbol{\theta}, \boldsymbol{\theta}_c, \boldsymbol{\theta}_p\}$ represents the comprehensive set of trainable parameters, encompassing the classical MLP weights ($\mathbf{W}^{(m)}$), the fusion rotation angles ($\boldsymbol{\theta}$), and the QCNN convolution ($\boldsymbol{\theta}_c$) and pooling ($\boldsymbol{\theta}_p$) parameters. For the $i$-th sample, $y_{k}^{(i)}$ is the one-hot encoded true label, and $\hat{y}_{k}^{(i)}$ is the model's predicted probability that the sample belongs to the $k$-th class. 

Parameters are optimized iteratively to minimize $\mathcal{L}(\boldsymbol{\Theta})$ using a gradient-based strategy. Gradients for the classical MLP are computed using standard backpropagation via the chain rule, while the parameters residing in the quantum circuit are updated using the parameter-shift rule. This method allows for exact gradient estimation directly on quantum hardware. Specifically, for a target quantum parameter $\phi_j$, the partial derivative of the class probability $\hat{y}_k$ is evaluated by shifting the parameter by macroscopic amounts:
\begin{equation}
\frac{\partial \hat{y}_k}{\partial \phi_j} = \frac{1}{2} \left( \hat{y}_k(\phi_j + \frac{\pi}{2}) - \hat{y}_k(\phi_j - \frac{\pi}{2}) \right)
\end{equation}
This value is subsequently incorporated into the chain rule to compute the final gradient with respect to the loss function:
\begin{equation}
\frac{\partial \mathcal{L}}{\partial \phi_j} = \sum_{k=0}^{C-1} \frac{\partial \mathcal{L}}{\partial \hat{y}_k} \frac{\partial \hat{y}_k}{\partial \phi_j}
\end{equation}
This formulation facilitates the seamless flow of gradients from the prediction output back through the quantum gates to the classical inputs, enabling unified end-to-end training.

\subsection{Theoretical Analysis and Properties}

\subsubsection{Parameter efficiency via exponential feature space} A fundamental advantage of the proposed architecture lies in the quantum circuit's ability to access an exponentially large feature space (Hilbert space) using only a linearly scaling number of parameters. We quantify this by analyzing the specific parameter complexity of the quantum component, denoted as $P_{q}$.

The trainable parameters within the quantum circuit consist solely of the rotation angles in the fusion layer and the variational parameters in QCNN. The fusion layer utilizes a bit-wise strategy requiring exactly one parameter $\theta_j$ per feature dimension. Thus the total number is $d$. QCNN comprises $L$ layers. The parameters for the convolution ($\boldsymbol{\theta}_c$) and pooling ($\boldsymbol{\theta}_p$) kernels are reused across the entire layer. Let $K = |\boldsymbol{\theta}_c| + |\boldsymbol{\theta}_p|$ denote the constant number of parameters within a single block (in our experiments, $K < 25$). In conclusion, the total parameter count for the quantum circuit is formulated as:
\begin{equation}
P_{q} = d + L \times K
\end{equation}
This formula demonstrates that the complexity of our quantum core scales linearly with the input dimension $d$ .

\subsubsection{Decomposability, parallelism, and trainability}
A critical bottleneck in scaling quantum neural networks is the barren plateau phenomenon, where the gradient variance decays exponentially with the total number of qubits, rendering the model untrainable. This issue typically stems from the global entanglement in deep, fully connected circuits.

Our architecture fundamentally overcomes this barrier through structural decomposability and parallelism. Due to the bit-wise independent topology of the fusion layer, the global quantum system is mathematically decoupled into a tensor product of unentangled local subsystems. Features at the corresponding positions  are independent of others during the fusion process. That is, both the forward state evolution and the backward gradient propagation are strictly confined within independent 3-qubit channels. The computationally expensive global operation is replaced by parallel, localized density matrix evolutions and partial traces. This implies that the calculation for a specific parameter is mathematically insulated from the noise or states of unrelated parallel qubits, effectively preventing the barren plateau and reducing the complexity of the model.

\subsubsection{Interpretability via evidence theory} Finally, we provide a formal mapping between the quantum fusion mechanics and DS theory. The fusion process is not an opaque operation but a realization of belief combination\cite{zhou2024transferable}.

We establish a formal behavioral isomorphism between the rule of combination and the quantum control mechanism. The control projection $|11\rangle\langle 11|$ in the $CC\text{-}R_y$ gate serves as a physical implementation of the \textit{conjunctive combination rule}. Let $\Omega$ be the frame of discernment. If we regard the active feature states as evidence sets $A \subseteq \Omega$ and $B \subseteq \Omega$ (respectively from two modals), the quantum activation condition corresponds strictly to the set-theoretic intersection $A \cap B$. Thus, the evolution of the target qubit is governed specifically by the joint consensus of the modalities, effectively mapping the logical conjunction of evidences onto the Hilbert space.

The rotation angle $\theta_j$ of fusion gate parameterizes the \textit{basic probability assignment}, denoted as $m(\cdot)$. The probability amplitude transfer can be modeled as assigning a belief mass to the fused feature: $m(\text{feature}_j) \propto \sin^2(\frac{\theta_j}{2})$. Thus, the trained parameter $\theta_j$ acts as a learnable reliability weight. A larger $\theta_j$ signifies that the model assigns higher belief mass to the joint evidence at index $j$, providing explicit transparency into the decision-making process.

\section{Experiments}
\label{sec:experiments}

\subsection{Dataset}

Houston2013\cite{debes2014hyperspectral} and Trento\cite{rasti2017fusion} two public specialized datasets, which are widely recognized and commonly used in the multimodal remote sensing field.

\textit{1) Houston2013:} This dataset was acquired over the University of Houston campus and its neighboring urban areas for the 2013 IEEE GRSS Data Fusion Contest. It consists of a hyperspectral image (HSI) and a co-registered LiDAR derived digital surface model, both possessing a spatial resolution of 2.5 m and an image size of 349 $\times$ 1905 pixels. The HSI data contains 144 spectral bands covering the wavelength range from 380 nm to 1050 nm. The ground truth includes 15 land cover classes, representing a complex urban environment.

\textit{2) Trento:} This dataset was captured over a rural area south of the city of Trento, Italy. It comprises HSI data acquired by the AISA Eagle sensor and LiDAR data acquired by the Optech ALTM 3100EA sensor. This dataset features a finer spatial resolution of 1 m and dimensions of 166 $\times$ 600 pixels. The HSI component consists of 63 spectral bands ranging from 402.89 nm to 989.09 nm. The ground truth contains 6 distinct land cover classes relevant to rural settings.

For the specific 4-class classification task in this study, we select multiple representative classes from these datasets to construct the training and testing sets, ensuring a balanced distribution of samples for the quantum circuit simulation.

\begin{table}[h!]
	\centering
	\caption{Classification Objective Statistics for Houston2013 and Trento Datasets}
	\label{tab:classification_objective}
	\vspace{-8pt} 
	\captionsetup[subfloat]{font=scriptsize} 
	\small
	
	\setlength{\tabcolsep}{3pt}
	
	\subfloat[Houston2013\label{tab:houston_stats}]{
		\begin{tabular}{l p{2cm} ccc}
			\toprule
			\textbf{Class ID} & \textbf{Class Name} & \textbf{Train samples} & \textbf{Test samples} & \textbf{Total} \\
			\midrule
			1 & Healthy grass & 1001 & 250 & 1251 \\
			2 & Stressed grass & 1004 & 250 & 1254 \\
			3 & Synthetic grass & 558 & 139 & 697 \\
			4 & Trees & 996 & 248 & 1244 \\
			5 & Soil & 994 & 248 & 1242 \\
			6 & Water & 260 & 65 & 325 \\
			7 & Residential & 1015 & 253 & 1268 \\
			8 & Commercial & 996 & 248 & 1244 \\
			9 & Road & 1002 & 250 & 1252 \\
			10 & Highway & 982 & 245 & 1227 \\
			11 & Railway & 988 & 247 & 1235 \\
			12 & Parking lot 1 & 987 & 246 & 1233 \\
			13 & Parking lot 2 & 376 & 93 & 469 \\
			14 & Tennis court & 343 & 85 & 428 \\
			15 & Running track & 528 & 132 & 660 \\
			\midrule
			\textbf{Total} & & 12030 & 2999 & 15029 \\
			\bottomrule
		\end{tabular}
	}
	
	\vspace{0.2cm}
	
	\subfloat[Trento\label{tab:trento_stats}]{
		\begin{tabular}{l p{2cm} ccc}
			\toprule
			\textbf{Class ID} & \textbf{Class Name} & \textbf{Train samples} & \textbf{Test samples} & \textbf{Total} \\
			\midrule
			1 & Apple trees & 323 & 80 & 403 \\
			2 & Buildings & 232 & 58 & 290 \\
			3 & Ground & 38 & 9 & 47 \\
			4 & Wood & 730 & 182 & 912 \\
			5 & Vineyard & 840 & 210 & 1050 \\
			6 & Roads & 254 & 63 & 317 \\
			\midrule
			\textbf{Total} & & 2417 & 602 & 3019 \\
			\bottomrule
		\end{tabular}
	}
\end{table}

\subsection{Experimental Setting}

\subsubsection{Experimental configuration} This study utilizes PyTorch and Pennylane frameworks for model construction and training  on an x86 platform (NVIDIA GeForce RTX 3090, 24G). The hyperparameters used in this study are detailed in \cref{tab:hyperparameters}. Quantum circuits are constructed and trained with Pennylane. Due to quantum resource limitations and the simulation capability of computers, our experiment implemented the model with 8 qubits. Specifically, the prepared 24 qubits is split into 8 groups of 3 qubits and simulated through torch tensor operations. Then Pennylane is only used to encode the density matrix of the fused target registers. The random number seeds for all experiments were fixed as 998244353 via function \texttt{pytorch\_lightning.seed\_everything}. 

\begin{table}[h]
	\centering
	\caption{Hyperparameter Configuration}
	\label{tab:hyperparameters}
	\begin{tabular}{cccc}
		\toprule
		\textbf{Learning Rate} & \textbf{Batch Size} & \textbf{Epochs} & \textbf{Random Seed} \\
		\midrule
		$1 \times 10^{-3}$ & 16 & 25 & 998244353 \\
		\bottomrule
	\end{tabular}
\end{table}

\begin{table*}[!t]
	\centering
	\small
	\caption{Parameter Statistics for Baseline Models}
	\label{tab:param_stats}
	
	\begin{tabular}{
			@{\hspace{0.1cm}} l 
			@{\hspace{0.1cm}} c 
			@{\hspace{0.38cm}} c 
			@{\hspace{0.38cm}} c 
			@{\hspace{0.38cm}} c 
			@{\hspace{0.38cm}} c 
			@{\hspace{0.38cm}} c 
			@{\hspace{0.38cm}} c 
			@{\hspace{0.38cm}} c
		}
		\toprule
		\textbf{Metric} & \textbf{EndNet} & \textbf{CrossFus.} & \textbf{FusAtNet} & \textbf{MDL-Middle} & \textbf{Classic-Fus.} & \textbf{Circuit-Block} & \textbf{All-to-All} & \textbf{QCMM (Ours)} \\
		\midrule
		\textbf{Total Parameters} & 85k & 99k & 17,440k & 99k & 2.4k & \textbf{1.7k} & \textbf{1.7k} & \textbf{2.2k} \\
		\textbf{Fusion Parameters} & 17k & 42k & 9,192k & 42k & 0.136k & \textbf{0} & \textbf{0} & \textbf{8} \\ 
		\textbf{Fusion Gate Count} & - & - & - & - & - & 16 & 24 & \textbf{8} \\
		\bottomrule
	\end{tabular}
\end{table*}

\begin{table*}[!t]
	\centering
	\small  
	\caption{Baseline Modal Analysis on the Houston2013 Dataset}
	\label{tab:houston_results}
	
	\begin{tabularx}{\textwidth}{l*{12}{>{\centering\arraybackslash}X}}
		\toprule
		& \multicolumn{4}{c}{\textbf{Circuit-Block Fusion}} & \multicolumn{4}{c}{\textbf{All-to-all Fusion}} & \multicolumn{4}{c}{\textbf{QCMM Fusion (Ours)}} \\
		\cmidrule(lr){2-5} \cmidrule(lr){6-9} \cmidrule(lr){10-13}
		\textbf{Metric} & $\mathbf{U_{SO4}}$ & $\mathbf{U_{SU4}}$ & $\mathbf{U_{15}}$ & \textbf{Avg.} & $\mathbf{U_{SO4}}$ & $\mathbf{U_{SU4}}$ & $\mathbf{U_{15}}$ & \textbf{Avg.} & $\mathbf{U_{SO4}}$ & $\mathbf{U_{SU4}}$ & $\mathbf{U_{15}}$ & \textbf{Avg.} \\
		\midrule
		C1 & 0.9440 & 0.9559 & 0.8679 & 0.9226 & 0.8920 & 0.8679 & 0.9440 & 0.9013 & 0.9520 & 0.9399 & 0.9599 & \textbf{0.9506} \\
		C2 & 0.9919 & 0.9919 & 0.9919 & 0.9919 & 0.5839 & 1.0000 & 1.0000 & 0.8613 & 1.0000 & 1.0000 & 0.9919 & \textbf{0.9973} \\
		C3 & 1.0000 & 1.0000 & 1.0000 & 1.0000 & 1.0000 & 1.0000 & 1.0000 & 1.0000 & 1.0000 & 1.0000 & 1.0000 & \textbf{1.0000} \\
		C4 & 0.9879 & 0.9758 & 0.9838 & 0.9825 & 0.9879 & 1.0000 & 0.9919 & \textbf{0.9933} & 0.9879 & 0.9919 & 0.9879 & 0.9892 \\
		\midrule
		OA & 0.9785 & 0.9785 & 0.9571 & 0.9714 & 0.8387 & 0.9627 & 0.9819 & 0.9278 & 0.9819 & 0.9842 & 0.9830 & \textbf{0.9830} \\
		AA & 0.9809 & 0.9809 & 0.9611 & 0.9743 & 0.8569 & 0.9670 & 0.9839 & 0.9359 & 0.9839 & 0.9859 & 0.9849 & \textbf{0.9849} \\
		Kappa & 0.9710 & 0.9709 & 0.9419 & 0.9613 & 0.7816 & 0.9496 & 0.9755 & 0.9022 & 0.9755 & 0.9786 & 0.9770 & \textbf{0.9770} \\
		F1 & 0.9789 & 0.9809 & 0.9610 & 0.9736 & 0.8540 & 0.9655 & 0.9835 & 0.9343 & 0.9839 & 0.9859 & 0.9849 & \textbf{0.9849} \\
		\bottomrule
	\end{tabularx}
\end{table*}

\begin{figure*}[!t]
	\centering
	\includegraphics[width=1\textwidth]{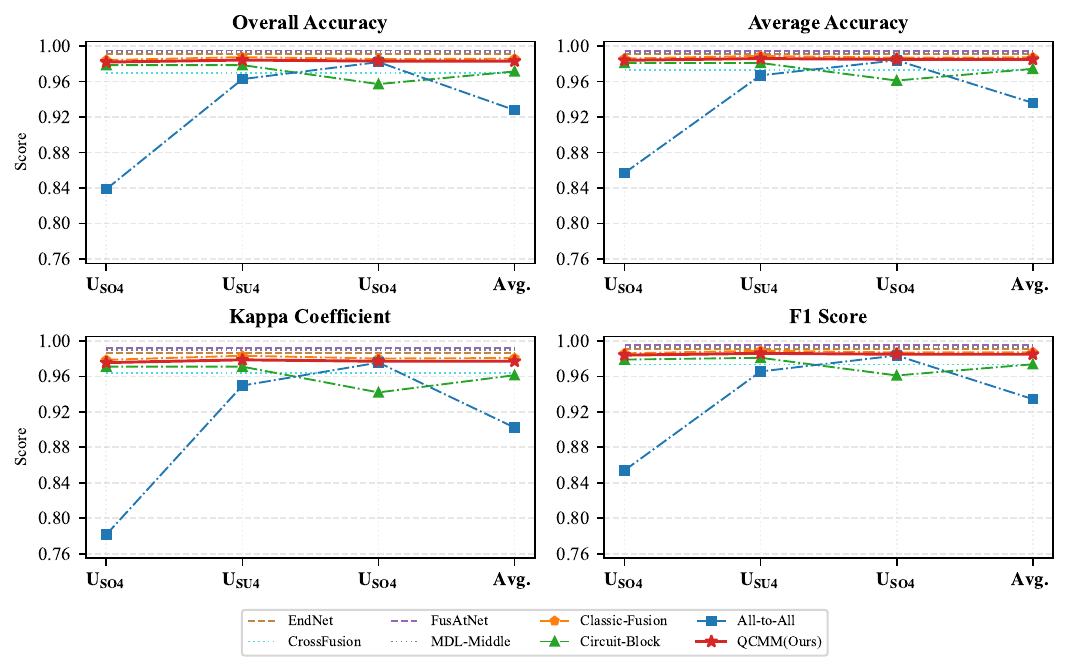}
	\caption{Comparison between classical models and quantum models. }
	\label{fig:comparison_result}
\end{figure*}

\subsubsection{Evaluation metrics} To comprehensively evaluate the classification performance of proposed QCMM, we employ four standard metrics in the remote sensing community: overall accuracy (OA), average accuracy (AA), the kappa coefficient ($\kappa$), and the F1-Score.

\subsection{Baseline Modal}
To comprehensively evaluate performance of the proposed QCMM, we benchmark it against a series of both quantum and classical baseline models. These models share similar preprocessing and feature extraction stages. All these experiments are conducted on the Houston2013.

\subsubsection{Comparison of quantum fusion strategies} To validate the effectiveness of our DS-Theory-based fusion strategy, we compare it against a classic fusion architecture two other prominent quantum fusion architectures as we illustrated in \hyperlink{target:fusion-strategy}{section II}. These fusion layers will all pass through same-hidden-layer unimodal MLP, and then connect the QCNN:

\textbf{QCMM (Ours)}: Implements the interpretable, bit-wise CC-$Ry(\theta)$ fusion mechanism as described in \hyperlink{target:fusion}{section III}.
	
\textbf{Circuit-Block}: Implements the structured entanglement pattern using CNOT gates with a fixed stride. 
	
\textbf{All-to-All}: A densely connected architecture where every HSI qubit is entangled with every LiDAR qubit.
	
\textbf{Classical-Fusion}: Two 8-dimensional vectors from both modalities are concatenated and sent into a single-layer MLP to obtain 8-dimensional fused data. This method has hundreds of parameters and is uninterpretable.

\subsubsection{Comparison with classical networks}To benchmark our quantum model against purely classical deep learning frameworks, we select four representative models that have demonstrated strong performance on remote sensing fusion tasks:

\textbf{EndNet}: An encoder-decoder architecture that utilizes a reconstruction strategy for feature fusion.

\textbf{CrossFusion / MDL-Middle}\cite{hong2020more}: These models based on a two-branch CNN structure, use weight sharing (CrossFusion) or intermediate concatenation (MDL-Middle) to achieve cross-modal interaction.

\textbf{FusAtNet}\cite{mohla2020fusatnet}: A dual-attention-based network that leverages a cross-attention mechanism to weight HSI features using LiDAR information.

\begin{table*}[!t]
	\centering
	\small 
	\caption{Ablation Study Analysis on the Houston2013 Dataset}
	\label{tab:ablation_study}
	
	\begin{tabularx}{\textwidth}{l*{12}{>{\centering\arraybackslash}X}}
		\toprule
		& \multicolumn{4}{c}{\textbf{w/o MLP}} & \multicolumn{4}{c}{\textbf{Fixed Fusion (CC-NOT)}} & \multicolumn{4}{c}{\textbf{Shallow QCNN}} \\
		\cmidrule(lr){2-5} \cmidrule(lr){6-9} \cmidrule(lr){10-13}
		\textbf{Metric} & $\mathbf{U_{SO4}}$ & $\mathbf{U_{SU4}}$ & $\mathbf{U_{15}}$ & \textbf{Avg.} & $\mathbf{U_{SO4}}$ & $\mathbf{U_{SU4}}$ & $\mathbf{U_{15}}$ & \textbf{Avg.} & $\mathbf{U_{SO4}}$ & $\mathbf{U_{SU4}}$ & $\mathbf{U_{15}}$ & \textbf{Avg.} \\
		\midrule
		C1 & 0.8479 & 0.8439 & 0.7279 & 0.8066 & 0.8719 & 0.8960 & 0.8799 & 0.8826 & 0.9399 & 0.9319 & 0.9480 & 0.9399 \\
		C2 & 0.6639 & 0.8719 & 0.5279 & 0.6879 & 1.0000 & 0.9959 & 1.0000 & \textbf{0.9986} & 0.9959 & 1.0000 & 1.0000 & \textbf{0.9986} \\
		C3 & 0.0647 & 0.0504 & 0.0504 & 0.0552 & 1.0000 & 1.0000 & 1.0000 & \textbf{1.0000} & 1.0000 & 1.0000 & 1.0000 & \textbf{1.0000} \\
		C4 & 0.6895 & 0.8911 & 0.9758 & 0.8521 & 0.9879 & 0.9879 & 0.9879 & 0.9879 & 0.9758 & 0.9758 & 0.9798 & 0.9771 \\
		\midrule
		OA & 0.6290 & 0.7452 & 0.6347 & 0.6696 & 0.9605 & 0.9661 & 0.9627 & 0.9631 & 0.9751 & 0.9740 & 0.9797 & 0.9763 \\
		AA & 0.5665 & 0.6659 & 0.5705 & 0.6010 & 0.9649 & 0.9699 & 0.9666 & 0.9671 & 0.9779 & 0.9769 & 0.9819 & 0.9789 \\
		Kappa & 0.4855 & 0.6458 & 0.4935 & 0.5416 & 0.9465 & 0.9541 & 0.9496 & 0.9501 & 0.9664 & 0.9648 & 0.9725 & 0.9679 \\
		F1 & 0.5423 & 0.6221 & 0.5377 & 0.5674 & 0.9645 & 0.9696 & 0.9666 & 0.9669 & 0.9766 & 0.9765 & 0.9819 & 0.9783 \\
		\bottomrule
		\\[-0.5em] 
		
		\toprule
		& \multicolumn{4}{c}{\textbf{HSI-only}} & \multicolumn{4}{c}{\textbf{LiDAR-only}} & \multicolumn{4}{c}{\textbf{QCMM (Ours)}} \\
		\cmidrule(lr){2-5} \cmidrule(lr){6-9} \cmidrule(lr){10-13}
		\textbf{Metric} & $\mathbf{U_{SO4}}$ & $\mathbf{U_{SU4}}$ & $\mathbf{U_{15}}$ & \textbf{Avg.} & $\mathbf{U_{SO4}}$ & $\mathbf{U_{SU4}}$ & $\mathbf{U_{15}}$ & \textbf{Avg.} & $\mathbf{U_{SO4}}$ & $\mathbf{U_{SU4}}$ & $\mathbf{U_{15}}$ & \textbf{Avg.} \\
		\midrule
		C1 & 0.9480 & 0.9480 & 0.8880 & 0.9280 & 0.8460 & 0.8560 & 0.8240 & 0.8420 & 0.9520 & 0.9399 & 0.9599 & \textbf{0.9506} \\
		C2 & 1.0000 & 0.9959 & 1.0000 & \textbf{0.9986} & 0.4600 & 0.4880 & 0.5519 & 0.5000 & 1.0000 & 1.0000 & 0.9919 & 0.9973 \\
		C3 & 1.0000 & 0.9785 & 1.0000 & 0.9928 & 1.0000 & 1.0000 & 1.0000 & \textbf{1.0000} & 1.0000 & 1.0000 & 1.0000 & \textbf{1.0000} \\
		C4 & 0.9798 & 0.9798 & 0.9879 & 0.9825 & 0.9959 & 0.9959 & 0.9879 & \textbf{0.9932} & 0.9879 & 0.9919 & 0.9879 & 0.9892 \\
		\midrule
		OA & 0.9797 & 0.9785 & 0.9650 & 0.9744 & 0.8083 & 0.8139 & 0.8207 & 0.8143 & 0.9819 & 0.9842 & 0.9830 & \textbf{0.9830} \\
		AA & 0.9814 & 0.9819 & 0.9689 & 0.9774 & 0.8299 & 0.8349 & 0.8409 & 0.8352 & 0.9839 & 0.9859 & 0.9849 & \textbf{0.9849} \\
		Kappa & 0.9725 & 0.9709 & 0.9526 & 0.9653 & 0.7422 & 0.7496 & 0.7589 & 0.7502 & 0.9755 & 0.9786 & 0.9770 & \textbf{0.9770} \\
		F1 & 0.9814 & 0.9819 & 0.9687 & 0.9773 & 0.8052 & 0.8132 & 0.8210 & 0.8131 & 0.9839 & 0.9859 & 0.9849 & \textbf{0.9849} \\
		\bottomrule
	\end{tabularx}
\end{table*}

\subsection{Ablation Study}
To rigorously validate the effectiveness of each component within the proposed QCMM framework, we designed three sets of ablation experiments. These experiments systematically dismantle key modules of the network to quantify their individual contributions to the final classification performance. These ablation experiments are set up as follows:

\textbf{Ablation of unimodal MLP layer (w/o MLP)}: We remove the trainable MLP responsible for unimodal feature extraction and alignment. The PCA-reduced classical vectors are directly fed into the quantum embedding layer.

\textbf{Ablation of fusion parameters (fixed fusion)}: In this experiment, the trainable rotation angles $\boldsymbol{\theta}_{f}$ in the CC-$Ry(\theta)$ fusion gates are fixed to $\theta = \pi$.

\textbf{Ablation of multimodal fusion (unimodal baselines)}: The model is trained using only HSI data or LiDAR data, processed through its dedicated MLP and the full QCNN module (no fusion layer).

\textbf{Ablation of QCNN depth (shallow QCNN)}{}: We reduce the depth of the QCNN by removing the second convolutional-pooling block. The QCNN will thus consist of only one block.

\subsection{Generalization Test}
\begin{figure*}[!t]
	\centering
	\captionsetup[subfloat]{font=scriptsize}
	
	\begin{minipage}{\textwidth}
		\centering
		\begin{minipage}{0.46\textwidth}
			\centering
			\subfloat[$\mathbf{U_5}$ ansatz]{
				\begin{adjustbox}{width=\linewidth}
					\begin{quantikz}[row sep=0.2cm, column sep=0.2cm]
						\lstick{} & \gate[style={rounded corners}]{R_x(\theta_1)} & \gate[style={rounded corners}]{R_z(\theta_3)} & \gate[style={rounded corners}]{R_z(\theta_5)} & \ctrl{1} & \gate[style={rounded corners}]{R_x(\theta_7)} & \gate[style={rounded corners}]{R_z(\theta_9)} & \qw \\
						\lstick{} & \gate[style={rounded corners}]{R_x(\theta_2)} & \gate[style={rounded corners}]{R_z(\theta_4)} & \ctrl{-1} & \gate[style={rounded corners}]{R_z(\theta_6)} & \gate[style={rounded corners}]{R_x(\theta_8)} & \gate[style={rounded corners}]{R_z(\theta_{10})} & \qw
					\end{quantikz}
				\end{adjustbox}
			}
		\end{minipage}%
		\hfill
		\begin{minipage}{0.145\textwidth}
			\centering
			\subfloat[$\mathbf{U_{TTN}}$ ansatz]{
				\begin{adjustbox}{width=\linewidth}
					\begin{quantikz}[row sep=0.2cm, column sep=0.3cm]
						\lstick{} & \qw & \gate[style={rounded corners}]{R_y(\theta_1)} & \ctrl{1} & \qw \\
						\lstick{} & \qw & \gate[style={rounded corners}]{R_y(\theta_2)} & \targ{}  & \qw
					\end{quantikz}
				\end{adjustbox}
			}
		\end{minipage}%
		\hfill
		\begin{minipage}{0.326\textwidth}
			\centering
			\subfloat[$\mathbf{U_{13}}$ ansatz]{
				\begin{adjustbox}{width=\linewidth}
					\begin{quantikz}[row sep=0.2cm, column sep=0.2cm]
						\lstick{} & \gate[style={rounded corners}]{R_y(\theta_1)} & \gate[style={rounded corners}]{R_z(\theta_3)} & \gate[style={rounded corners}]{R_y(\theta_4)} & \ctrl{1} & \qw \\
						\lstick{} & \gate[style={rounded corners}]{R_y(\theta_2)} & \ctrl{-1}  & \gate[style={rounded corners}]{R_y(\theta_5)} & \gate[style={rounded corners}]{R_z(\theta_6)} & \qw
					\end{quantikz}
				\end{adjustbox}
			}
		\end{minipage}
	\end{minipage}
	
	\vspace{0.2cm} 
	
	\begin{minipage}{\textwidth}
		\centering
		\begin{minipage}{0.46\textwidth}
			\centering
			\subfloat[$\mathbf{U_6}$ ansatz]{
				\begin{adjustbox}{width=\linewidth}
					\begin{quantikz}[row sep=0.2cm, column sep=0.2cm]
						\lstick{} & \gate[style={rounded corners}]{R_x(\theta_1)} & \gate[style={rounded corners}]{R_z(\theta_3)} & \gate[style={rounded corners}]{R_x(\theta_5)} & \ctrl{1} & \gate[style={rounded corners}]{R_x(\theta_7)} & \gate[style={rounded corners}]{R_z(\theta_9)} & \qw \\
						\lstick{} & \gate[style={rounded corners}]{R_x(\theta_2)} & \gate[style={rounded corners}]{R_z(\theta_4)} & \ctrl{-1} & \gate[style={rounded corners}]{R_x(\theta_6)} & \gate[style={rounded corners}]{R_x(\theta_8)} & \gate[style={rounded corners}]{R_z(\theta_{10})} & \qw
					\end{quantikz}
				\end{adjustbox}
			}
		\end{minipage}%
		\hfill
		\begin{minipage}{0.173\textwidth}
			\centering
			\subfloat[$\mathbf{U_9}$ ansatz]{
				\begin{adjustbox}{width=\linewidth}
					\begin{quantikz}[row sep=0.2cm, column sep=0.3cm]
						\lstick{} & \gate[style={rounded corners}]{H} & \ctrl{1} & \gate[style={rounded corners}]{R_x(\theta_1)} & \qw \\
						\lstick{} & \gate[style={rounded corners}]{H} & \ctrl{-1} & \gate[style={rounded corners}]{R_x(\theta_2)} & \qw
					\end{quantikz}
				\end{adjustbox}
			}
		\end{minipage}%
		\hfill
		\begin{minipage}{0.326\textwidth}
			\centering
			\subfloat[$\mathbf{U_{14}}$ ansatz]{
				\begin{adjustbox}{width=\linewidth}
					\begin{quantikz}[row sep=0.2cm, column sep=0.2cm]
						\lstick{} & \gate[style={rounded corners}]{R_y(\theta_1)} & \gate[style={rounded corners}]{R_x(\theta_3)} & \gate[style={rounded corners}]{R_y(\theta_4)} & \ctrl{1} & \qw \\
						\lstick{} & \gate[style={rounded corners}]{R_y(\theta_2)} & \ctrl{-1}  & \gate[style={rounded corners}]{R_y(\theta_5)} & \gate[style={rounded corners}]{R_x(\theta_6)} & \qw
					\end{quantikz}
				\end{adjustbox}
			}
		\end{minipage}
	\end{minipage}
	
	\caption{Architectures of additional ansatzes for generalization test.}
	\label{fig:additional_kernels}
\end{figure*}
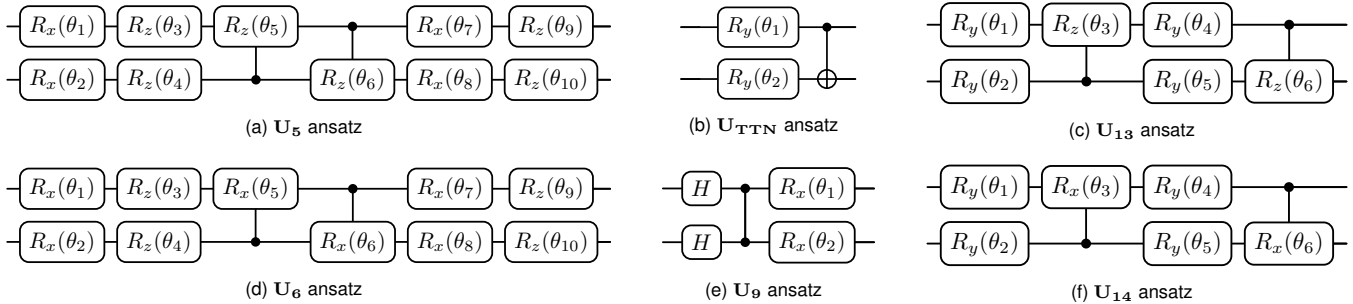

We construct multiple distinct classification tasks by selecting different subsets of land cover classes from both datasets Houston2013 and Trento. This setup tests the model's adaptability. Since addressing domain shifts and data distribution differences across varying scenes is a critical challenge in remote sensing \cite{dong2025multimodal}, validating the model's robustness on these diverse subsets is essential. For Houston2013, we design four distinct subsets covering all the classes: Group A $\{1, 2, 3, 4\}$, Group B $\{5, 6, 7, 8\}$, Group C $\{8, 9, 10, 11\}$, and Group D $\{12, 13, 14, 15\}$. For Trento, we design three subsets: Group E $\{1, 2, 3, 4\}$, Group F $\{1, 2, 5, 6\}$, and Group G $\{3, 4, 5, 6\}$. The categories represented by different labels can be corresponding in \cref{tab:classification_objective}. Besides, we add more convolution kernels summarized by Hur et al. \cite{hur2022quantum} for testing as fig. \ref{fig:additional_kernels} shows. This demonstrates the robustness of our structure.

\begin{table*}[!t]
	\centering
	\small
	\caption{Generalization Test Results (OA) on Houston2013 and Trento Datasets}
	\label{tab:generalization_test}
	
	\begin{tabularx}{\textwidth}{l>{\raggedright\arraybackslash}X*{9}{>{\centering\arraybackslash}X}}
		\toprule
		\textbf{Dataset} & \textbf{Group} & $\mathbf{U_{TTN}}$ & $\mathbf{U_{5}}$ & $\mathbf{U_{6}}$ & $\mathbf{U_{9}}$ & $\mathbf{U_{13}}$ & $\mathbf{U_{14}}$ & $\mathbf{U_{15}}$ & $\mathbf{U_{SO4}}$ & $\mathbf{U_{SU4}}$ \\
		\midrule
		\multirow{4}{*}{Houston2013} & A & 0.9775 & 0.9808 & 0.9808 & 0.9786 & 0.9797 & 0.9820 & 0.9830 & 0.9819 & \textbf{0.9842} \\
		& B & 0.9496 & 0.9459 & 0.9496 & 0.9226 & 0.9472 & \textbf{0.9607} & 0.9447 & 0.9509 & 0.9570 \\
		& C & 0.8859 & 0.8737 & 0.8626 & 0.8727 & 0.8758 & 0.8707 & \textbf{0.8899} & 0.8667 & 0.8838 \\
		& D & 0.9586 & \textbf{0.9676} & 0.9622 & 0.9245 & 0.9532 & 0.9568 & 0.9550 & 0.9604 & 0.9514 \\
		\midrule
		\multirow{3}{*}{Trento} & E & 0.9939 & 0.9878 & 0.9939 & 0.9970 & 0.9970 & \textbf{1.0000} & 0.9970 & 0.9939 & 0.9909 \\
		& F & 0.9538 & 0.9611 & 0.9538 & 0.9416 & 0.9489 & 0.9489 & 0.9635 & 0.9635 & \textbf{0.9659} \\
		& G & \textbf{0.9957} & 0.9935 & 0.9935 & 0.9763 & 0.9784 & 0.9784 & \textbf{0.9957} & 0.9871 & 0.9914 \\
		\bottomrule
	\end{tabularx}
\end{table*}

\subsection{Performance Analysis}

\subsubsection{Comparison with quantum fusion strategies} To evaluate the ability and the robustness of our DS-theory-based fusion strategy, we compared QCMM against other quantum baselines (\textit{Circuit-Block}, \textit{All-to-All}, \textit{Classic-Fusion}) across 4 different efficient quantum convolutional kernels.

The result shows that our fusion method has the highest accuracy among quantum fusion methods. In the tests of the three highest-performing convolution kernels, the accuracy rates reached 0.9819, 0.9842, and 0.9830 respectively.  And it's only 0.3\% worse than the classic fusion method when we only have 8 fusion parameters while the classic  fully connected fusion layer has hundreds.

While strategies like \textit{Circuit-Block} and  \textit{All-to-All} suffer from noise accumulation in deep circuits, our proposed fusion mechanism effectively balances expressibility and trainability. This demonstrates that the bit-wise entangled fusion, guided by evidence theory, provides a robust framework that is general to the choice of the downstream convolutional kernel.

\subsubsection{Comparison with classical models} We benchmarked the proposed QCMM against several classical deep learning models, including EndNet, CrossFusion, FusAtNet, and MDL-Middle. OA of these four models are 0.9910, 0.9700, 0.9950 and 0.9930 respectively. The result illustrates QCMM can maintain competitive classification accuracy with a significant advantage in parameter efficiency and interpretability.

QCMM outperforms the CrossFusion. Furthermore, compared to the best-performing classical model, our model achieves an accuracy gap within 1\%, demonstrating its capability to capture complex spatial-spectral features effectively. The most notable advantage lies in the model complexity. QCMM requires significantly fewer trainable parameters—approximately 1/40 of those used in EndNet or CrossFusion, and merely 1/7900 of the parameters in FusAtNet. This massive reduction (by orders of magnitude) confirms that QCMM can achieve state-of-the-art performance with minimal computational resources, validating the power of quantum entanglement in feature compression and representation.

\subsubsection{Ablation study analysis} The ablation experiments further validate the necessity of each component in the QCMM framework:

\textbf{Effect of unimodal MLP layer}: Removing the unimodal MLPs resulted in a notable performance drop, the accuracy of the three highest-performing convolution kernels' tests drops to 0.6290, 0.7452 and 0.6347, confirming their critical role in unimodal feature extraction and implicitly aligning the semantic features of HSI and LiDAR for effective quantum fusion.

\textbf{Effect of trainable bit-wise fusion}: Fixing the fusion parameters to $\pi$ (degrading equivalent to CC-NOT gates) reduced accuracy. The accuracy drops to  0.9605, 0.9661 and 0.9627. This proves that the trainable rotation angles $\theta$ successfully capture the importance weights (belief mass) of different features, confirming the value of our evidence-theory-based design.

\textbf{Effect of multimodal fusion}: The QCMM significantly outperforms both HSI-only and LiDAR-only baselines that only have the accuracy of 0.9797, 0.9785, 0.9650 (HSI) and 0.8083, 0.8139, 0.8207 (LiDAR), verifying that the model successfully leverages the complementary information from both modalities.

\textbf{Effect of QCNN layers}: With only one set of convolution and pooling, the accuracy drops to 0.9751, 0.9740 and 0.9797. The performance degradation in the shallow QCNN variant confirms that the hierarchical convolution-pooling structure is essential for abstracting high-level semantic features from the fused quantum state.

\subsubsection{Generalization test} The experimental results indicate that QCMM maintains consistent high performance across all tested subsets. Despite the significant differences in spectral signatures and spatial structures among these groups, the model achieved stable accuracy with minimal fluctuation. This evidence strongly suggests that the QCMM does not merely memorize specific class attributes but successfully learns generic, discriminative spatial-spectral representations. The proposed quantum fusion and feature extraction mechanisms exhibit strong generalization potential, making the model adaptable to diverse remote sensing classification scenarios.

\section{Conclusion}
\label{sec:conclusion}

In this article, we have introduced QCMM, a novel quantum multimodal fusion framework for multimodal remote sensing multi-classification tasks. Our model consists of three parts: the classical unimodal feature extraction aligner, the quantum multimodal fusion layer, and QCNN. Its innovative fusion method, as a decomposable structure, effectively reduces computational complexity, ensuring scalability for high-dimensional data while providing the interpretability of evidence. The model exhibited consistent stability and high accuracy across nine different quantum convolution kernels.

For future work, firstly, we will leverage the high scalability of QCMM to explore the fusion of additional modalities. Secondly, we will delve deeper into the interpretability of the fusion parameters. Visualizing trained rotation angles in the fusion gates with feature may provide novel insights into the model's decision-making process. And explore more interpretable fusion frameworks.


\bibliographystyle{IEEEtran}
\bibliography{refs}

\vfill

\end{document}